\documentclass[twocolumn,times,tighten]{aastex62}

\shorttitle{Late-forming dwarfs in $\Lambda$CDM}
\shortauthors{Benitez-Llambay \& Fumagalli}

\defcitealias{Benitez-Llambay2020}{BLF}

\begin{document}

\title{The Tail of Late-forming Dwarf Galaxies in $\Lambda$CDM}

\correspondingauthor{Alejandro Benitez-Llambay}
\email{alejandro.benitezllambay@unimib.it}

\author[0000-0001-8261-2796]{Alejandro Benitez-Llambay}
\author[0000-0001-6676-3842]{Michele Fumagalli}
\affil{Dipartimento di Fisica G. Occhialini, Universit\`a degli Studi di Milano Bicocca, Piazza della Scienza, 3 I-20126 Milano MI, Italy}

\begin{abstract}
We use a robust analytical model together with a high-resolution hydrodynamical cosmological simulation to demonstrate that in a $\Lambda$ cold dark matter ($\Lambda$CDM) universe, a small fraction of dwarf galaxies inhabiting dark matter (DM) halos in the mass range $3\times 10^{9} \lesssim M_{200} / M_{\odot} \lesssim 10^{10}$ form unusually late ($z<3$) compared to the bulk population of galaxies. These galaxies originate from the interplay between the stochastic growth of DM halos and the existence of a time-dependent DM halo mass below which galaxies do not form. The formation epoch of the simulated late-forming galaxies traces remarkably well the time when their host DM halos first exceeded a nontrivial (but well-understood) time-dependent critical mass, thus making late-forming dwarfs attractive cosmological probes with constraining power over the past growth history of their host halos. The agreement between our model and the simulation results demonstrates that the population of simulated late-forming dwarfs is a robust cosmological outcome and largely independent of the specific galaxy formation model included in the simulations provided: (1) the universe underwent cosmic reionization before $z_{\rm re} \sim 8$; (2) star formation proceeds in gas that self-gravitates; and (3) galaxy formation is largely restricted to atomic-cooling halos before $z_{\rm re}$. The scarcity of massive late-forming dwarfs expected in $\Lambda$CDM implies that the great majority of bright, metal-poor, and actively star-forming dwarfs observed in our local universe--the most obvious candidates for these late-forming galaxies--cannot be undergoing their formation for the first time at the present day in a $\Lambda$CDM universe.
\end{abstract}

 
\section{Introduction} \label{sec:intro}
Within the Lambda cold dark matter ($\Lambda$CDM) cosmological model, galaxies form from gas that collapses in the center of gravitationally bound dark matter (DM) halos~\citep{White1978}. Galaxies do not form, however, in every halo. In the absence of external heating sources, galaxy formation is restricted to the so-called atomic-cooling (AC) halos, i.e., halos that shock heat the gas to a temperature $T \gtrsim 10^{4} \rm \ K$, above which radiative cooling becomes efficient. Under the presence of external energetic sources that suppress atomic cooling and heat the gas~\cite[e.g.][]{Efstathiou1992}, such as the ultraviolet background (UVB) radiation field that keeps the universe (re)ionized, galaxy formation proceeds in halos that exceed a time-dependent critical mass above which the gas becomes self-gravitating. The value of this critical mass has been estimated in the past using either idealized Jeans arguments or numerical simulations~\cite[e.g.][]{Rees1986, Thoul1996, Quinn1996},\footnote{Note that the critical mass that results from simulations is usually expressed as a Jeans mass, or equivalently as a fix halo circular velocity cut below which galaxies do not form.} leading to the understanding that for galaxies to form after cosmic reionization (CR), their host halos must exceed the AC limit by a factor of a few. The AC limit is, therefore, useful to determine which halos host galaxies at high redshift, before the universe underwent CR~\citep[e.g.][]{Oh2002, Xu2013, Wise2014, Xu2016}, whereas the critical mass describes the onset of galaxy formation after CR. The onset of galaxy formation is thus deeply linked to the growth of DM halos, as only those halos that exceed the critical mass can host a luminous galaxy in their center. These ideas have been shown to agree qualitatively with results of hydrodynamical cosmological simulations~\cite[e.g.][and references therein]{Okamoto2009, Sawala2016, Benitez-Llambay2015, Benitez-Llambay2017, Fitts2017, Maccio2017}, and are of fundamental importance to explain the scarcity of observed luminous satellites compared to results of collisionless simulations~\citep{Klypin1999, Bullock2000}. However, it is clear that neither halos have uniform density nor the interstellar medium is isothermal, so the successful characterization of the critical mass necessarily requires more advanced modeling.~\citet[]{Benitez-Llambay2020} (hereafter BLF) have recently derived the critical mass for the onset of galaxy formation considering the nonlinear collapse of DM halos and the distinctive temperature-density relation of the intergalactic medium, removing the freedom inherent to Jeans arguments. Their critical mass (hereafter BLF mass), which differs from that arising from idealized Jeans modeling, is remarkably accurate compared to results of nonlinear hydrodynamical cosmological simulations.

The existence of a critical mass for galaxy formation, coupled with the growth history of DM halos, has some interesting consequences. First, galaxies must form stochastically in halos of present-day mass under the critical mass~(\citetalias{Benitez-Llambay2020}). Second, galaxies residing in more massive halos today must form, on average, earlier than those inhabiting less massive halos, giving rise to a ``downsizing'' effect in a $\Lambda$CDM universe~\cite[e.g.][]{Neistein2006}. Finally, there must be a population of galaxies residing in DM halos with a present-day mass comparable to critical mass that has undergone their formation only recently. Robustly quantifying these expectations is, however, not trivial, as it requires precise knowledge of the assembly history of DM halos and the critical mass for galaxy formation.

In this Letter, we address the last issue. In particular, we apply the recent~\citetalias{Benitez-Llambay2020} model (briefly described in Sec.~\ref{Sec:model}) together with a high-resolution hydrodynamical cosmological simulation to demonstrate that the existence of a population of late-forming dwarf galaxies that form after redshift $z=3$ is a robust cosmological outcome of the $\Lambda$CDM model, and highly insensitive to the simulation details, provided stars form in gas that self-gravitates. This prediction stems from the existence of a nontrivial time-dependent critical DM halo mass below which galaxy formation cannot take place and the stochastic growth that characterizes DM halos in $\Lambda$CDM, which generally depends on the cosmological parameters~\cite[e.g.][and references therein]{Lacey1993, VanDenBosch2002, Correa2015}. We quantify the abundance of late-forming dwarfs expected in a $\Lambda$CDM universe, and possible observational counterparts, in Sec.~\ref{Sec:Discussion}.

\section{The Model and Simulation}

\subsection{The BLF model} \label{Sec:model}
The~\citetalias{Benitez-Llambay2020} model establishes the value of the critical mass for the onset of galaxy formation as a function of time. In this model, galaxy formation takes place in AC halos before CR, and in halos in which gas cannot remain in hydrostatic equilibrium afterward. To determine which halos undergo gravitational collapse to form a galaxy after CR, the model assumes that the gas that falls into dark halos is described by an effective equation of state, which is established by the photoheating background at low densities ($n_{\rm H} \lesssim 10^{-4.5} \rm \ cm^{-3}$), and by the interplay between photoheating and cooling at high densities. This nontrivial model thus avoids the common assumption that galaxy formation takes place in halos of given (fix) virial\footnote{We define virial quantities as those measured at the virial radius, $r_{200}$, defined as the radius of a sphere whose mean enclosed density is 200 times the critical density of the universe.} temperature, a condition that arises from idealized Jeans arguments.\footnote{The condition that galaxies form in halos for which the freefall time exceeds the sound-crossing time leads to the well-known condition for galaxy formation, $V_{\rm c} >> c_{s}$, in which $c_{s}$ is the sound speed of an ``isothermal'' intergalactic medium, and $V_{\rm c}$ is the halo circular velocity at $r_{\rm 200}$.} We refer the reader to the original \citetalias{Benitez-Llambay2020} paper for further details and a derivation of the critical mass.

\begin{figure}
    \centering
    \includegraphics{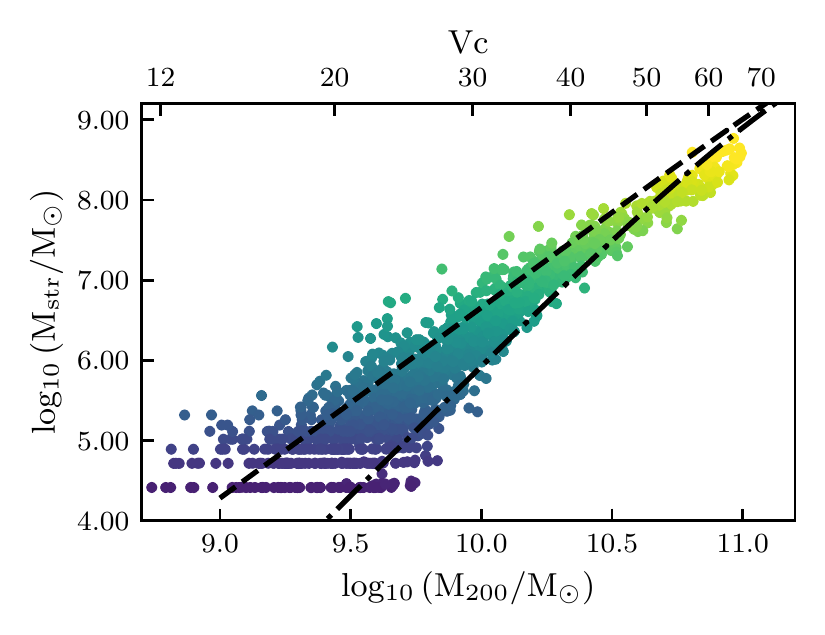}
    \caption{Stellar vs. halo mass relation for our sample of simulated ``central'' galaxies. Dotted-dashed and dashed lines show abundance-matching expectations from~\cite{Guo2010} and~\cite{Moster2013}, respectively. Galaxies are colored according to their current stellar mass ($y$-axis).}
    \label{Fig:Mstr_M200}
\end{figure}

\subsection{The Simulation}
We use a high-resolution hydrodynamical cosmological simulation carried out with the {\tt P-Gadget 3} code~\citep[last described in][]{Springel2005} along with the {\tt EAGLE} model of galaxy formation~\citep{Schaye2015}. The simulation is the same introduced in~\cite{Benitez-Llambay2020} and we list here only the physical prescriptions relevant for our work. We refer the reader to the original papers for further details. The simulation evolves a periodic cubic volume of side length $20 \rm \ Mpc$, filled with $2\times 1024^3$ DM and gas particles, so that the DM and gas particle mass are $m_{\rm dm} = 2.9 \times 10^{5} \ M_{\odot}$ and $m_{\rm gas} = 5.4 \times 10^{4} \ M_{\odot}$, respectively. The adopted Plummer-equivalent gravitational softening is $\epsilon=195 \rm \ pc$. Gas particles are turned into collisionless star particles once they exceed a density threshold of $n_{\rm thr} = 1 \rm \ cm^{-3}$, a sufficiently high value that ensures that the gas within DM halos becomes self-gravitating before turning into stars. Unlike the original {\tt EAGLE} simulations, our density threshold for star formation does not depend on metallicity. Our simulation also includes radiative cooling and heating, as tabulated by~\citet{Wiersma2009}, which in turn includes the~\citet{Haardt2001} UVB radiation field. CR is modeled by turning on the UVB at redshift $z_{\rm re}=11.5$ and to ensure that gas is quickly heated to $\sim 10^{4} \ K$ at $z_{\rm re}$, an energy of $2 \rm \ eV$ per proton mass is instantaneously injected to every gas particle at that time. 

DM halos are identified in the simulation using {\tt HBT+}~\citep{Han2018}, which provides a catalog of ``central'' and ``satellite'' DM halos identified in the simulation using a friends-of-friends algorithm with percolation length $b=0.2$ in units of the mean interparticle separation, and assigning ``bound'' particles to each halo based on their binding energies.

\subsection{Sample Selection}
We select simulated galaxies as ``central'' DM halos that contain more than one stellar particle within their galactic radius, $r_{\rm gal} = 0.2 \times r_{200}$, which yields a lower galaxy stellar mass limit of $\sim 5.4 \times 10^{4} \ M_{\odot}$. As shown in Fig.~\ref{Fig:Mstr_M200}, this criterion imposes a minimum present-day halo mass $M_{200} \sim 10^{9} \ M_{\odot}$. We also restrict our sample to DM halos with virial mass $M_{200} \lesssim 10^{11} M_{\odot}$, as objects above this limit are not of interest for our study because all these systems exceed the AC limit prior to CR~\citep{Benitez-Llambay2020}, thus making it impossible for these halos to host galaxies that form late. In the selected mass range, the stellar mass of our simulated dwarfs are bounded from above and below by the abundance-matching (AM) constraints from~\citet{Moster2013} and~\citet{Guo2010}, respectively (see Fig.~\ref{Fig:Mstr_M200}). At large halo masses, the mass of the central galaxies are underestimated in our simulation compared to AM expectations, a limitation of the original {\tt EAGLE} model that does not preclude the analysis that follows.

\section{Results}

\subsection{The Simulated Tail of Late-forming Dwarfs}

In order to look for a population of late-forming galaxies, we shall consider the galaxy formation time, $t_{\rm f}$, defined as the time at which a galaxy first formed a star particle in the simulation. In practice, $t_{\rm f}$ corresponds to the formation time of the oldest star particle found within $r_{\rm gal}$ at $z=0$.

\begin{figure}
    \centering
    \includegraphics{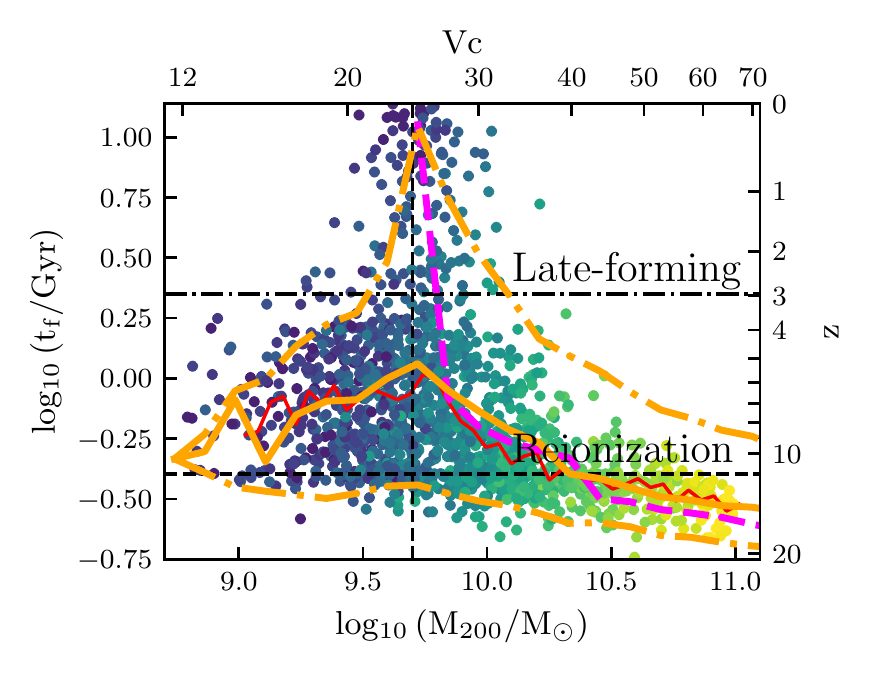}
    \caption{Galaxy formation time, $t_{\rm f}$, as a function of the present-day host halo mass (symbols are colored as in Fig.~\ref{Fig:Mstr_M200}). The thin red solid line indicates the running median of the distribution; the thick magenta dashed line shows the time when a mean EPS halo of mass $M_{200}$ first exceeds the~\citetalias{Benitez-Llambay2020} mass. Orange lines are similar to the thick dashed line, but they show the median and the 10th and the 90th percentiles that result from comparing individual (rather than average) EPS halo growth histories to the~\citetalias{Benitez-Llambay2020} mass. See the text for further discussion.}
    \label{Fig:formation_time}
\end{figure}

Fig.~\ref{Fig:formation_time} shows $t_{\rm f}$ as a function of present-day virial mass for our galaxy sample. Galaxies inhabiting more massive halos at redshift $z=0$ form earlier than those hosted by less massive counterparts, although the median $t_{\rm f}$ (shown by the thin red solid line) is weakly dependent on present-day halo mass. More interesting is the fact that the scatter in $t_{\rm f}$ increases significantly at low masses, peaking at about the present-day value of the~\citetalias{Benitez-Llambay2020} mass (vertical dashed line). The large scatter at low masses originates from a tail of late-forming dwarfs that we arbitrarily define as galaxies with $t_{\rm f} \gtrsim 2.2 \rm \ Gyr$ (or $z \lesssim 3$), and that constitutes less than $\sim 8 \%$ of population of dwarfs with stellar masses $M_{\rm gal} \gtrsim 5.4 \times 10^{5} \ M_{\odot}$ in the halo mass range $3 \times 10^{9} \lesssim M_{200} / M_{\odot} \lesssim 10^{10}$. But what determines, in general, the $t_{\rm f}$ versus $M_{200}$ relation shown in Fig.~\ref{Fig:formation_time}, and more particularly, what is the origin of the tail of late-forming dwarfs identified in our simulation? 

\begin{figure}
    \centering
    \includegraphics{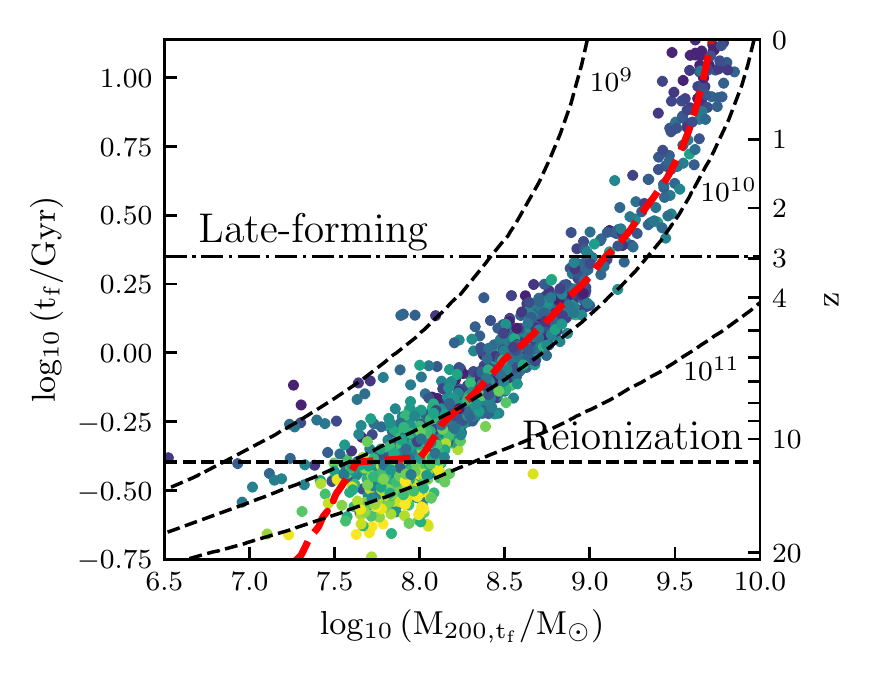}
    \caption{Galaxy formation time, $t_{\rm f}$, as a function of the halo virial mass at $t_{\rm f}$. Galaxies are colored as in Fig.~\ref{Fig:Mstr_M200}. The red dashed line shows the time-dependent critical mass for galaxy formation that results from the~\citetalias{Benitez-Llambay2020} model (see Sec.~\ref{Sec:model}). The critical mass exhibits a sudden increase from $M_{200} \sim 10^{7.5} \  M_{\odot}$ to $M_{200} \sim 10^{8.0} \  M_{\odot}$ at CR. Dashed lines show the $\Lambda$CDM mean mass growth histories calculated using the EPS formalism, for halos of present-day virial mass $10^{9}, 10^{10}$, and $10^{11} \ M_{\odot}$.}
    \label{Fig:critical_mass}
\end{figure}

In light of the discussion of Sec.~\ref{sec:intro}, one may speculate that the main trend of the $t_{\rm f}$ versus $M_{200}$ relation displayed by our galaxy sample is related to the interplay between the mean assembly history of their host halos and the evolution of the critical mass for galaxy formation. We demonstrate that this is indeed the case in Fig.~\ref{Fig:critical_mass}, which shows $t_{\rm f}$ as a function of halo mass, but as measured at the galaxy formation time rather than at $z=0$, $M_{200,t_{\rm f}}$. We compare the values measured in the simulation to the time-dependent ~\citetalias{Benitez-Llambay2020} mass (red dashed line). This critical mass displays a sharp transition from $M_{\rm crit} \sim 3 \times 10^{7} \ M_{\odot}$ to $M_{\rm crit} \sim 10^{8} \ M_{\odot}$ at $z=z_{\rm re}$, and results from the fact that galaxy formation is largely limited by atomic hydrogen cooling before CR, and by the ability of the gas to undergo gravitational collapse afterward.
Fig.~\ref{Fig:critical_mass} makes it clear that galaxy formation occurs predominantly in halos whose virial mass at $t_{\rm f}$ is comparable to the critical mass. Interestingly, the~\citetalias{Benitez-Llambay2020} mass traces the simulated $t_{\rm f}$ versus $M_{200,t_{\rm f}}$ relation remarkably well, even though neither the~\citetalias{Benitez-Llambay2020} model nor the {\tt EAGLE} model were tuned to do so. 

\subsection{Origin of the Overall Mass Dependence of Galaxy Formation Time}

The black dashed lines in Fig.~\ref{Fig:critical_mass} show three mean $\Lambda$CDM mass growth histories derived using the Extended Press-Schechter (EPS) formalism~\cite[][]{Bond1991}, for halos of present-day mass, $10^{11}$, $10^{10}$ and $10^{9} \ M_{\odot}$. These curves help to clarify the overall mass dependence of the galaxy formation time observed in Fig.~\ref{Fig:formation_time}. Consider, for example, the upper dashed curve, which corresponds to the mean mass growth of a halo with present-day mass, $M_{200}=10^{9} \ M_{\odot}$. This demonstrates that the bulk population of $10^{9} M_{\odot}$ halos cannot host a galaxy in their center, as the mean mass growth history for these halos is under the~\citetalias{Benitez-Llambay2020} mass at all times. On the other hand, massive halos today, those that greatly exceed the present-day critical mass for galaxy formation, have exceeded the~\citetalias{Benitez-Llambay2020} mass since before CR. Consequently, these halos host galaxies that have formed very early on, explaining to some extent the observed ``downsizing'', as argued in Sec.~\ref{sec:intro}. The typical formation time for those intermediate-mass halos that eventually exceed the critical mass depends monotonically on halo mass.

These arguments indicate that it is possible, in principle, to understand the overall mass-dependent galaxy formation time. To this end, we must compare the mean mass growth of $\Lambda$CDM halos to the time-dependent critical mass for galaxy formation, and define $t_{\rm f}$ as the time when the mean halo mass first exceeds the~\citetalias{Benitez-Llambay2020} mass. The thick magenta dashed line in Fig.~\ref{Fig:formation_time} shows the result of this exercise. The agreement between this curve and the median $t_{\rm f}$ that results from the simulation (thin solid line), particularly at high masses, demonstrates that our interpretation is correct. However, this simple exercise results in a divergent $t_{\rm f}$ toward the present-day value of the~\citetalias{Benitez-Llambay2020} mass (vertical dashed line), as no galaxy formation can take place below this mass scale, as anticipated. This reasoning would imply that all galaxies inhabiting halos with present-day mass $M_{200} \sim M_{\rm crit,0}$ are young, contrary to the simulation results, in which only a low fraction of the systems form at late times. It is thus evident that although the mean mass growth of $\Lambda$CDM halos is useful to understand the overall trend observed in Fig.~\ref{Fig:formation_time}, it is not sufficient to account for the predominantly old population of galaxies inhabiting halos with present-day mass $M_{200} \lesssim M_{\rm crit,0}$. The existence of these galaxies in our simulation is a direct consequence of the fact that galaxy formation becomes an increasingly rare event at low halo masses, so the proper modeling of the galaxy formation time at these masses must necessarily take into account the intrinsic scatter in the growth of $\Lambda$CDM halos.  

\subsection{Origin of the Scatter in the Galaxy Formation Time}

To account for the scatter in the mass growth of halos, we now consider individual (rather than average) EPS growth histories in bins of present-day halo mass in the range, $10^{9} \lesssim M_{200} / M_{\odot} \lesssim 10^{11}$. We sample the scatter in the assembly of halos by constructing 500 growth histories per mass bin. As in the previous section, we define $t_{\rm f}$ as the time at which the EPS mass first crosses the~\citetalias{Benitez-Llambay2020} mass. The thick orange lines of Fig.~\ref{Fig:formation_time} show the result of this calculation, in particular the median and the 10th and 90th percentiles of the distributions. By considering the intrinsic scatter in the growth history of halos we eliminate the divergence of $t_{\rm f}$ toward the present-day value of the critical mass (vertical line). This is because the great majority of these halos formed their galaxies at earlier times, albeit at much later times than massive halos. This improved model naturally reproduces the increasingly large scatter in $t_{\rm f}$ at low masses. The peak in the 90th percentile around the critical mass is in remarkable agreement with the results from the simulation, so we can safely conclude that the mass-dependent galaxy formation time is simply understood from the interplay between the critical mass for galaxy formation and the time when the halos first exceeded this mass. Within this picture, an unavoidable conclusion is that in $\Lambda$CDM there must be a low fraction of halos that have exceeded this critical mass for galaxy formation particularly late ($z<3$), thus triggering the first episode of star formation at late times. Late-forming dwarfs can thus be regarded as robust cosmological outcomes, being their formation time largely insensitive to the details of the modeling included in numerical simulations.\footnote{This is true provided star formation proceeds in gas that self-gravitates, a requirement that arises from the definition of the critical mass in the~\citetalias{Benitez-Llambay2020} model.}

Finally, DM halos less massive than the critical mass today host predominantly old galaxies, similar to more massive halos. As opposed to late-forming dwarfs, these galaxies populate the low fraction of halos that collapsed unusually early to become more massive than the critical mass at earlier times~\citep[see][for a detailed discussion on this]{Benitez-Llambay2020}. Some of these halos would host the observed population of ancient ultrafaint dwarfs~\cite[e.g.][]{Simon2019}, which we do not resolve in our simulation.

\begin{figure}
    \centering
    \includegraphics{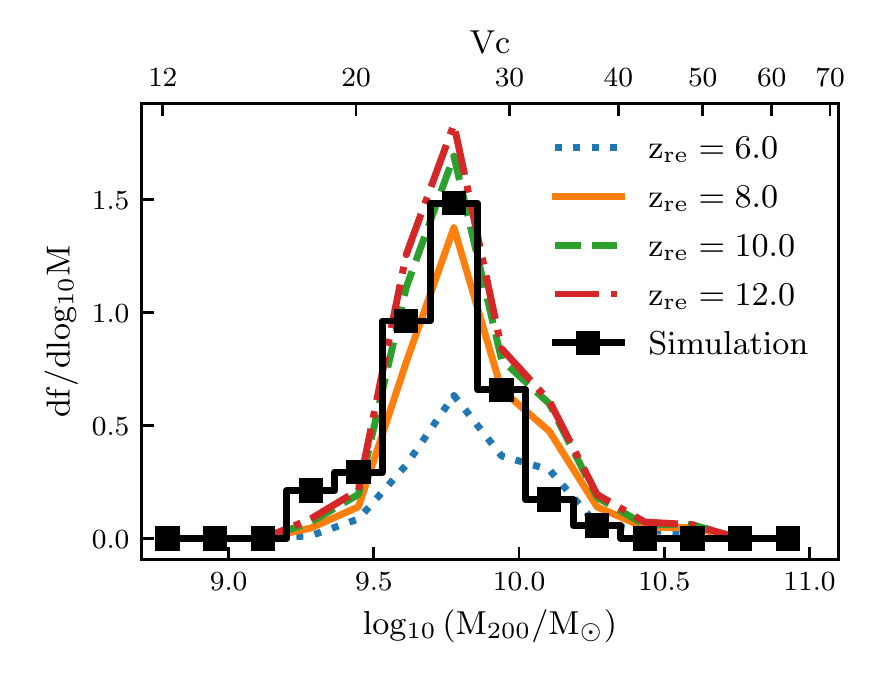}
    \caption{Frequency of late-forming galaxies, i.e., those that formed after redshift $z\lesssim 3$, as a function of halo mass. The black histogram shows the the result of our simulation. The other lines display the expected fraction of late-forming dwarfs for different values of the redshift of reionization, $z_{\rm re}$, as indicated in the legend. If reionization occurs earlier than $z_{\rm}=8$, our predictions become insensitive to the particular value of $z_{\rm ze}$.}
    \label{Fig:impact_reionization}
\end{figure}

\section{Discussion and Conclusions} \label{Sec:Discussion}

Our analysis demonstrates that the existence of a small population of late-forming galaxies, here defined as dwarfs that started forming stars after $z<3$, is a robust outcome of the $\Lambda$CDM model. This is because their formation depends on the ability of the gas to undergo gravitational collapse, which does not depend on the particularities of the galaxy formation model assumed in simulations. Indeed, Fig.~\ref{Fig:critical_mass} demonstrates that the galaxy formation time is a fair measure of the time when the $\Lambda$CDM halos first exceeded the critical mass above which gas cannot remain in hydrostatic equilibrium. Moreover, we showed in Fig.~\ref{Fig:formation_time} that the~\citetalias{Benitez-Llambay2020} model, together with the EPS formalism, enables us to understand not only the dependence of the galaxy formation time on halo mass but also its scatter. 

Our results are also robust to the assumed redshift of reionization. Fig.~\ref{Fig:impact_reionization} shows the frequency of late-forming dwarfs as measured in our simulation and as obtained from comparing EPS mass growth histories to the~\citetalias{Benitez-Llambay2020} mass, assuming that the universe undergoes reionization at $z_{\rm re}=6$, $8$, $10$, and $12$. Clearly, the frequency of late-forming dwarfs is largely insensitive to the exact value of $z_{\rm re}$ provided $z_{\rm re} \gtrsim 8$, a lower limit consistent with recent Planck results~\citep{Planck2020}.

Finally, we note that our results rely on the idea that galaxy formation largely proceeds in AC halos prior to cosmic reionization~\citep[e.g.,][and references therein]{Bromm2011}. Numerical simulations that include physical ingredients missed in our simulation and that are important to address this issue --such as molecular hydrogen cooling and radiative feedback effects-- support the idea that the first galaxies indeed form predominantly in atomic cooling halos~\cite[e.g.][and references therein]{Oh2002, Greif2008, Wise2014}. In particular, the work by~\cite{Wise2014} shows that the fraction of halos that host galaxies prior to cosmic reionization vanishes in a narrow range of halo mass below the AC limit. Our results thus rest on assumptions that appear plausible and warrant further scrutiny. 

\subsection{Late-forming Dwarfs in Other Simulations}

Interestingly, late-forming dwarfs have already been spotted in cosmological hydrodynamical simulations. For example, using a high-resolution zoom-in simulation of the formation of the Local Group from the {\tt CLUES} project,~\cite{Benitez-Llambay2015} identified two dwarf galaxies with a significant delay in their formation, and they ascribed their origin to the impact of CR. Similarly, using a sample of 15 zoom-in cosmological simulations carried out with the {\tt FIRE} code,~\cite{Fitts2017} identified one dwarf galaxy that formed after $z<1$. They, too, ascribed the unusual delayed formation of this galaxy to the effect of CR. Although not discussed by the authors, a small fraction of the isolated low-mass dwarf galaxies analyzed by~\cite{Garrison-Kimmel2019} also formed particularly late compared to most dwarfs in their simulations. Finally, recent works have shown that dwarf galaxies that form early can undergo late episodes of star formation once their halos experience significant mergers, a phenomenon related to the fact that DM halos can exceed the critical mass more than once during their lifetime~\cite[e.g.][and references therein]{Benitez-Llambay2016, Rey2020}. This suggests that the critical mass for galaxy formation does not only establish the onset of galaxy formation for starless halos, but also determines the ability of luminous halos to collect sufficient gas to sustain star formation in their center, and may shape the star formation history of dwarfs.\footnote{The gas in DM halos less massive than the critical mass is stable against gravitational collapse and therefore unable to form stars~\citep{Benitez-Llambay2020}, provided the stellar content of the halo is gravitationally irrelevant.}

All these results thus strongly support the idea that late-forming dwarfs are rare objects that arise naturally in numerical simulations, regardless of the adopted modeling. 

\subsection{Predicted Number Density}

\begin{figure}
    \centering
    \includegraphics{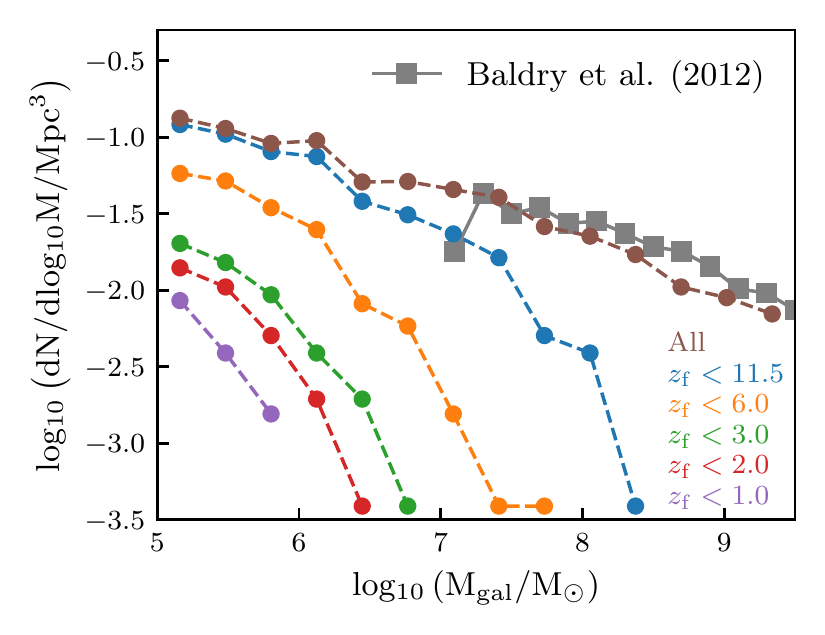}
    \caption{Stellar mass function of the simulated galaxies (brown dashed line), split by the galaxy formation redshift, $z_{\rm f}$, as indicated in the legend. Late-forming dwarfs (defined here as those with $z_{\rm f} < 3$), make up less than $20\%$ of the simulated galaxy population at the low-mass end, and they contribute even less at larger stellar masses.}
    \label{Fig:number_density}
\end{figure}

The number density of late-forming dwarfs depends on the abundance of halos that exceed the~\citetalias{Benitez-Llambay2020} mass after $z\lesssim 3$. In Fig.~\ref{Fig:number_density} we show the galaxy stellar mass function of the simulated galaxy population, and of galaxies that formed their first stars after redshift $z_{\rm f}$. As anticipated throughout our Letter, only a low fraction of galaxies make up the population of late-forming dwarfs (i.e., those with $z_{\rm f} < 3$). Moreover, the fraction of late-forming dwarfs decreases steadily with decreasing formation redshift and with increasing stellar mass. Finding a massive dwarf ($M_{\rm gal} > 10^{7} \ M_{\odot}$)  undergoing its formation today becomes thus extremely rare; in fact, dwarfs undergoing their formation at the present day are expected to be faint, with masses comparable to the faint and ultrafaint dwarfs observed nearby ($M_{\rm gal} \lesssim 10^{6} \ M_{\odot}$). Although this particular statement might depend on the particular galaxy formation model included in our simulation, the good match between the observed galaxy stellar mass function from~\cite{Baldry2012} and that measured in our simulation (brown line in Fig.~\ref{Fig:number_density}) suggests that the stellar masses of the simulated dwarfs are robust, at least in the mass range of overlap.

\subsection{Observational Counterparts of the Late-forming Dwarfs}

It is natural to expect that late-forming galaxies exhibit systematically smaller sizes than the majority of dwarfs. This is because late-forming dwarfs form from gas that dissipates its thermal energy and sinks to the center at late times, preventing their young stars from being stirred up by mergers, as opposed to older dwarfs~\cite[see, e.g.,][for a discussion of this effect in dwarf galaxies]{Benitez-Llambay2016}. Our simulation indicates that the half-mass radius of late-forming dwarfs resolved with more than 50 particles is indeed, on average, $30 \%$ smaller than that of older dwarfs of similar present-day stellar mass. Besides their compactness, these galaxies form from generally pristine gas. Therefore, the most obvious candidates for the $\Lambda$CDM late-forming dwarfs are some of the most metal-poor star-forming blue compact dwarfs (BCDs) known in the Local Volume. BCDs such as I Zwicky 18 or DDO 68 are indeed characterized by compact sizes, ongoing intense star-formation activity in their center, and surprisingly low metallicity. Due to the difficulty of reconstructing star-formation histories with high resolution at early times, it is however unclear whether they contain a substantial population of old stars~\cite[e.g.][]{Papaderos2002, Izotov2004, Pustilnik2005, Pustilnik2008, Jamet2010}.

Consider, for example, the case of I Zwicky 18. The vigorous star formation at its center at a rate of $\sim 16 \times 10^{-2} \ M_{\odot} \rm \ yr^{-1} \ kpc^{-2}$, together with its large gas supplies, low dust abundance~\cite[e.g.][]{Wu2007}, low-metallicity of $\sim 1/50 \  Z_{\odot}$~\cite[e.g.][]{Aloisi1999}, and slow circular velocity of $(38 \pm 4.4) \rm \ km \ s^{-1}$~\citep{Lelli2012}, make this galaxy an ideal analog for the most massive galaxies found in our sample of simulated late-forming dwarfs whose gaseous halo has started the runaway gravitational collapse not so long ago. Also, the extreme properties of I Zwicky 18 have led some authors to argue that this dwarf would be a present-day analog of star-forming galaxies found at high redshift~\cite[e.g.][]{Papaderos2012}. However, the idea that I Zwicky 18 is a dwarf undergoing its first formation today has been disputed on the grounds that, similarly to other observed BCDs~\cite[e.g.][]{Schulte-Ladbeck2001, Annibali2003,  Vallenari2005, Aloisi2005, Makarov2017}, I Zwicky 18 contains RGB stars that put a lower limit constraint to its formation time of about 1 Gyr ago~\citep[e.g.][and references therein]{Aloisi2007, Annibali2013, Sacchi2016}. This limit, albeit old in the context of stellar population synthesis models, only constitutes a small fraction of the cosmic scales spanned by the late-forming dwarfs discussed here. 

On statistical grounds, however, the scarcity of massive late-forming dwarfs in $\Lambda$CDM indicates that most BCDs observed in the local universe cannot be undergoing their formation for the first time today. If our volume is representative of the local universe, our simulation suggests that the great majority of late-forming dwarfs should be much fainter than I Zwicky 18, and much fainter than most BCDs. Indeed, BCDs have stellar masses in the range $10^{7} \lesssim M_{\rm gal}/M_{\odot} \lesssim 10^{9}$, whereas Fig.~\ref{Fig:number_density} shows that simulated late-forming dwarfs have stellar masses $M_{\rm gal} \lesssim 10^{7} \ M_{\odot}$. Whether one of the known BCDs with extreme properties constitutes a true analog of the late-forming dwarfs analyzed here could be answered with deeper photometric observations that reach the oldest main-sequence turnoff~\cite[see, e.g., the review by][]{Gallart2005}. 

The higher resolution and sensitivity afforded by upcoming observational facilities, in particular the James Webb Space Telescope and the Extremely Large Telescope, will enable precise measurements of resolved star formation histories in these types of galaxies at larger distances, helping to constrain their formation epoch and placing them in the context of the population analyzed in our work. On the other hand, upcoming surveys such as those by the Vera Rubin Observatory may be able to detect ultrafaint dwarfs beyond our Local Group, some of which may well be the late-forming dwarfs discussed here. These endeavors should not be taken lightly, as late-forming dwarfs could provide a powerful avenue to elucidate the past growth of low-mass DM halos, probe a distinctive halo mass-scale, and test the core of our understanding of galaxy formation at the smallest scales.

\newpage 
\acknowledgments

We thank the anonymous referee for a thoughtful review of our manuscript, which led to the improvement of our presentation. This work is supported by the European Research Council (ERC) under the European Union's Horizon 2020 research and innovation program (GA 757535) and UNIMIB's Fondo di Ateneo Quota Competitiva (project 2020-ATESP-0133).
This work used the DiRAC Data Centric system at Durham University, operated by the Institute for Computational Cosmology on behalf of the STFC DiRAC HPC Facility (https://www.dirac.ac.uk). This equipment was funded by BIS National E-infrastructure capital grants ST/P002293/1, ST/R002371/1, and ST/S002502/1, Durham University, and
STFC operations grant ST/R000832/1. DiRAC is part of the National e-Infrastructure. The simulation used in this work was performed using DiRAC’s director discretionary time awarded to A.B.L. A.B.L. is grateful to Prof. Mark Wilkinson for awarding this time.

\bibliography{bibliography}

\end{document}